\documentstyle[aps,prd,eqsecnum]{revtex}

\input psfig.sty

\begin{document}
\draft
\title{The Copernican Principle in Compact Spacetimes}
\author{John D. Barrow and Janna Levin}
\address{ DAMTP, Centre for Mathematical Sciences, Cambridge University,\\
Wilberforce Rd., Cambridge CB3 0WA }

\twocolumn[\hsize\textwidth\columnwidth\hsize\csname
           @twocolumnfalse\endcsname

\maketitle
\widetext

\begin{abstract}

Copernicus realised we were not at the centre of the
universe.  A universe made finite by topological identifications
introduces a new Copernican consideration: while we may not be at the
geometric centre of the universe, some galaxy could be.  A finite
universe also picks out a preferred frame: the frame in which the
universe is smallest. Although we are not likely to be at the centre of the
universe, we must live in the preferred frame (if we are at rest with
respect to the cosmological expansion). We show that the
preferred topological frame must also be the comoving frame in a
homogeneous and isotropic cosmological spacetime.  Some implications
of topologically identifying time are also discussed.

\end{abstract}

\medskip
\noindent{}
\medskip
]

\narrowtext

\section{Introduction}

Copernicus asserted that the Sun rather than the Earth is at the centre of
the solar system. Cosmologists have generalised this Copernican Principle
into a form of  Cosmological Principle: that our place in the universe is
not special within the set of habitable locations \cite{bt}. A homogeneous
and isotropic Friedmann universe with simply-connected
topology is generally consistent with this cosmological expression of the
Copernican Principle since no spatial locations are preferred. However, a
topologically finite universe introduces a new ingredient to the issue of
special locations. A compact manifold will possess a geometric 
centre\footnote{
The origin can be defined as the maximum of the injectivity radius 
\cite{Thurston}. The only exceptions to this are the homogeneous compact spaces
such as the sphere $S^{3}$, the Projective Plane, and the hypertorus $T^{3}$. 
As a result of invariance under translations, any point in the manifold
could equally well be the origin.}. But the geometric centre is not in any
sense the location of the big bang. The big bang happened everywhere
simultaneously. The geometric centre is only the most symmetrically located
point in a compact space; for further discussion see the reviews on topology
and cosmology in refs. \cite{{lum},{cqg},{review}}.\ 

In addition to a geometric centre, there is also a preferred frame.
To see this, it is
helpful to first consider the situation in topologically identified
flat space. In a finite, flat spacetime, a variant of the twin paradox
arises which exposes the existence of a preferred frame 
\cite{{bl},{weeks}}. One twin stays on Earth. The other travels at
constant velocity out into space. They both believe the other's time dilates
so which one is actually younger when the space-travelling twin returns home
and they meet again? The twin paradox in an infinite Minkowski
spacetime is resolved by recognising the asymmetry that arises because
the space traveller must stop and turn around and therefore experience
accelerated motion. In a finite spacetime this resolution no longer exists.
The space-travelling twin can remain inertial and still return home by
travelling on a periodic geodesic. If they are both inertial, which one is
younger when the travelling twin passes by the stay-at-home twin? The space
travelling twin is again younger. The  paradox is resolved by the existence
of a preferred frame which is introduced by the finite topology. It
is the frame in which the universe is seen to be smallest. In this preferred
coordinate system, space is finite and clocks can be synchronized. For any
observer moving at constant velocity relative to this topologically
preferred frame, the finite space will appear distorted, spacetime points
will be identified with other spacetime points, and there is no way to
synchronize clocks consistently \cite{{bl},{weeks},{peters},{lh}}.

Instead of flat spacetime, consider an expanding homogeneous and isotropic
Friedmann-Robertson-Walker (FRW) universe. These spacetimes select
a different
preferred frame, namely the coordinate system in which
observers comove with the cosmological expansion. Only the comoving
observers are inertial since any initial peculiar velocity or rotational
velocity perturbation in the cosmological fluid will die away as the
universe expands so long as the pressure $p$ and density $\rho $ of the
fluid satisfy $p<\rho /3$ . This suggests that the only inertial observers
are comoving observers and so, if the space is both expanding and
topologically compact, the preferred topological frame must be the comoving
frame. This we will show in what follows.

\section{Compact Space and a Preferred Frame}

To demonstrate the existence of a preferred frame, we consider again the
compactification of a flat Minkowski spacetime. Let our coordinate system be 
$x=(\tau ,\vec{x})$. In a flat, static spacetime, 
\begin{equation}
ds^{2}=-d\tau ^{2}+d\vec{x}^{2}=-d\tau ^{\prime 2}+d\vec{x}^{\prime 2},
\end{equation}
where we have also introduced $x^{\prime }=(\tau ^{\prime },\vec{x}^{\prime
})$, the coordinate system of an observer travelling at constant velocity
relative to $x,$ so that 
\begin{equation}
x^{\prime }=\Lambda x
\end{equation}
with the Lorenz transformation 
\begin{equation}
\Lambda =\pmatrix{\cosh b & \sinh b\cr \sinh b & \cosh b }\,
\end{equation}
where $b$ is a boost related to the relative velocity by $v=\tanh b$. The
Lorenz transformation is a rotation in Minkowski spacetime. 

\begin{figure}[tbp]
\centerline{\psfig{file=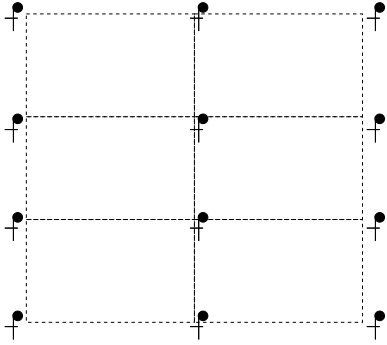,width=2.in}}
\vskip 5truept
\caption{A two-dimensional slice through a finite volume for the preferred
observer $x^\prime$. A baby is born at a well-defined time in the
hypertorus. There are an infinite number of repeated ghost images of the
birth.}
\label{prefer}
\end{figure}

Let space be topologically identified into a simple hypertorus in frame $x$.
For an observer in this frame, the topology only affects the spatial
coordinates and leaves the time coordinate unchanged. We represent the
topological boundary conditions by 
\begin{equation}
x\rightarrow \gamma x
\end{equation}
where $\gamma $ is an element of the discrete subgroup of the full group of
isometries, $\gamma \in \Gamma $. The compact manifold is ${\cal M}\sim 
{\cal M}^{U}/\Gamma $ where ${\cal M}^{U}$ is the simply-connected universal
cover -- in this case, flat spacetime. An observer with worldline $x$ will
interpret the finite space as a simple tiling of the universal cover as in
figure \ref{prefer}.

Suppose we are in relative motion with respect to the primed frame. The
topological identification appears to us as the condition 
\begin{equation}
x^{\prime }\rightarrow \Lambda \gamma x\,.  \label{bc}
\end{equation}
Suppressing all but one of the spatial dimensions for simplicity, we reduce
the hypertorus to a circle of circumference $L$. The identification is
effected by the simple translation $x\rightarrow \gamma x$ which is
equivalent to the condition that $(\tau ,x)\rightarrow (\tau ,x+L)$ and
eqn.\ (\ref{bc}) becomes\footnote{We could have written 
$\gamma $ as an explicit matrix by embedding our $(1+1) $-dimensional 
spacetime in a $(2+1)$-dimensional spacetime with the
additional spatial coordinate fixed \cite{{bl},{weeks}}. For this simple
flat spacetime it isn't necessary, although it is extremely useful when
working in curved spacetimes.} 
\begin{equation}
\pmatrix{\tau^\prime \cr x^\prime}\rightarrow \pmatrix{\tau^\prime +L\sinh b
\cr x^\prime + L \cosh b}\,.
\end{equation}
One point in space or time is identified with another point in space 
{\it and time}.

Notice that in the $x$-frame, the spatial size of the universe is $L$
while in the $x^\prime$-frame, the spatial size of the universe is 
$L\cosh b>L$. Of course the space{\it time} interval is invariant. However the 
universe appears smallest 
to the observer at rest with respect to the topological identification.
The topology thereby selects this preferred frame. It is also the frame in 
which clocks can be properly synchronized.

\begin{figure}[tbp]
\centerline{\psfig{file=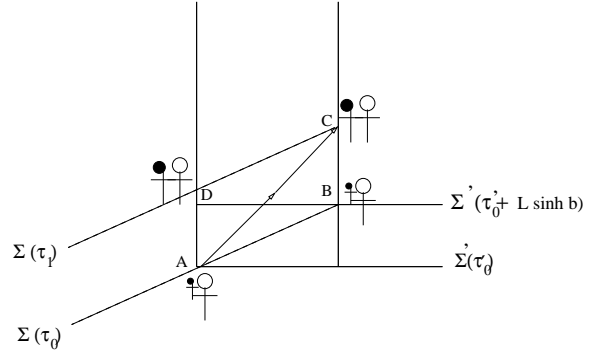,width=3.0in}}
\vskip 5truept
\caption{The spacetime diagram in coordinate system $x^\prime$. 
$\Sigma^\prime$ is the spacelike hypersurface foliated by 
$\protect\tau^\prime $. Also drawn for reference is $\Sigma$, the 
spacelike hypersurface
foliated by $\protect\tau$. A baby is born at spacetime point A:
($\protect\tau^\prime_0,x^\prime_0)$. The baby is also born at 
another spacetime 
point B: $(\protect\tau_0^\prime+ L\sinh b,x^\prime_0+ L\cosh b)$. A and B are
topologically identified. The mother travels from A to C but can only arrive
after the birth so she will never observe the paradox that the same baby was
born more than once. The points C and D are topologically identified. }
\label{frames}
\end{figure}

The spacetime diagram is drawn in figure \ref{frames}. Suppose a baby is
born at spacetime point $(\tau^\prime_{0},x^\prime_{0})$. 
The baby must also be born at 
$(\tau^\prime_{0}+L\sinh b,x_0^\prime+L\cosh b)$. 
That is, the baby also appears to be born
at another time and place - in fact, at an infinite number of other times
and places. Of course, the baby can only be born once, and all observers
must agree that the baby is only born once. For instance, as shown in figure 
\ref{frames}, if the mother of the child travelled near light speed from A
with the intention of arriving at B to relive the birth, she would
inevitably arrive late. The time it would take her to intercept the baby's
worldline is $\Delta x/\tanh b=L\cosh b/\tanh b>L\sinh b$ and
superluminal motion would be required to arrive in time. Still, she {\it is}
simultaneously at spacetime points A and B in figure \ref{frames}. What time
does she think it is? She might decide it is time $\tau _{0}^{\prime }$ or
she might decide it is time $\tau _{0}^{\prime }+L\sinh b$. Regardless, she
would find that it was impossible to synchronize her clocks 
\cite{{bl},{weeks},{peters},{lh}}. 
Despite the identification of one time with another, 
we will never experience a unique event twice.
For null and timelike observers, there is no way to travel back in time.

\section{Compact Space and the Comoving Frame}

Topology is of cosmological interest. We may very well live in a universe
that is finite. In a finite space there is definite meaning to the question,
Where are we in space? We can search the cosmic background
radiation (CMB) for evidence of the size, shape, and our location in space.
In fact, preliminary results from WMAP (Wilkinson Microwave Anisotropy Probe)
do give hints of a finite extent to
the universe \cite{{wmap},{angelica},{efstath}}.

For the purposes of searching for topology in CMB maps, the topological size
of the universe would have to be smaller than or just comparable to the
particle horizon to have observable consequences. Since the universe appears
to be homogeneous and isotropic within the observable horizon, research is
often restricted to finite spaces which support a FRW metric over the
entire space. There exists a frame in which observers at rest with respect
to that frame see topological identifications of spatial sections only.
Observers in motion relative to this frame will see a mixture of spatial and
temporal identifications.

\begin{figure}[tbp]
\centerline{\psfig{file=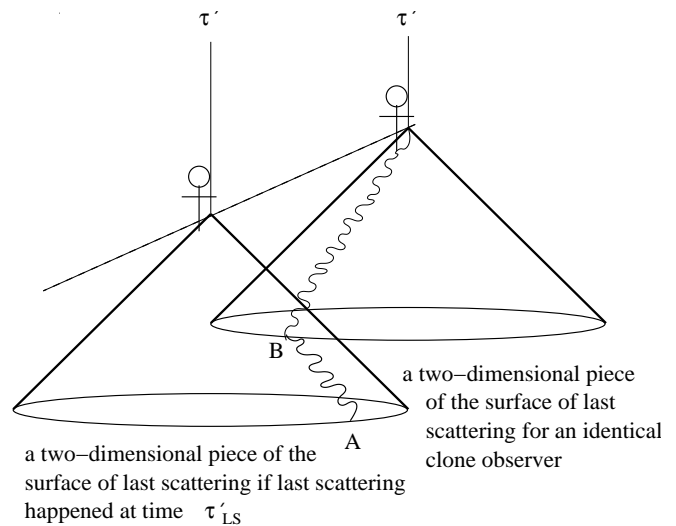,width=3.4in}}
\vskip 5truept
\caption{An observer in the $x^\prime$ frame 
looks back billions of years to the time
of last scattering and assumes last scattering happened at one well-defined
time $\protect\tau^\prime_{LS}$ only to find that this cannot be the case.
Notice that the surfaces of last scattering of two clone observers do not
intersect implying no circles and no repeated images in the light they
receive. However, light emitted at A can travel to hit
another point on the surface of last scatter at B, before B was emitted.
Therefore the emission of light from B could not be simultaneous with the
emission of light from A. Last scattering could not occur at a well defined
time in $x^\prime$. This is another manifestation of the 
observers' inability to synchronize their clocks in this frame.}
\label{SLS}
\end{figure}

Notice also that if $x^{\prime }$, which is not the preferred topological
frame, were also the comoving frame in an expanding universe, observers
would see no patterns and no circles \cite{{css},{pat}} in cosmic microwave
background (CMB) maps. All directions in the sky would give a unique point
on a CMB map, as illustrated in figure \ref{SLS}.

However, if the universe is expanding, then $x$ and not $x^{\prime }$ must
be the comoving frame. For if $x^{\prime }$ were the comoving frame it
would follow that 
\begin{equation}
ds^{2}=a(\tau ^{\prime })^{2}\left( -d\tau ^{^{\prime }2}+d\vec{x}^{^{\prime
}2}\right) ,
\end{equation}
with $a(\tau ^{\prime })$ the homogeneous and isotropic FRW scale factor.
The scale factor would have to match the boundary condition 
\begin{equation}
a(\tau ^{\prime })=a(\tau ^{\prime }+L\sinh b),
\end{equation}
which is impossible if $a(\tau ^{\prime })$ evolves monotonically. By
contrast, if $x$ is the comoving frame then there are no additional
constraints on $a(\tau )$ since compactification occurs on a purely
spacelike hypersurface. In other words, unless the comoving frame is
coincident with the preferred topological frame, the local Einstein
equations will not be satisfied. Still, no galaxy is truly at rest with
respect to the cosmic expansion. One might wonder if any peculiar motion
with respect to the comoving frame
would degrade a detection of topology through patterns and circles.

This suggests that however the universe may have begun, if there is
initially a preferred topological frame the expansion must settle down to
become the comoving frame. 
The topology of the 3-large spatial dimensions, along with the
topology of any extra small dimensions, is fixed in the very early universe
presumably by quantum gravitational processes. Initially the cosmos may have
been very inhomogeneous and anisotropic, even chaotic. Some physical process
drives the universe to the homogeneous and isotropic state we observe.
(Interestingly, we know that the most general
homogeneous anisotropic spaces which provide the modes which render the 
negatively curved
FRW models unstable are excluded by finite topological identifications 
\cite{{kod1},{kod2}}.)\footnote{These 
considerations suggest that finite topology
can play some role in providing a basis for Mach's Principle \cite{king}.} 
Baring inflation\footnote{If the universe inflates, 
then the topological size of the large dimensions
would naturally be well beyond the observable horizon. In which case, the
universe is not homogeneous and isotropic across the entire manifold.}, if
the universe really is smooth across the entire space, then one can argue
heuristically that a physical smoothing mechanism should align the preferred
topological frame with the comoving frame. Any smoothing that requires
causal communication across the entire manifold will be restricted to
communicate via modes on that manifold which are consistent with the
boundary conditions. Only in the preferred topological frame will modes
settle into a statistically
homogeneous and isotropic form, without shearing. That frame
will therefore naturally settle down to be the comoving frame.

\section{Compact Time}

In the simplest global FRW model of the universe it is not possible to make
time compact since that would require $a(\tau )=a(\tau +nT)$ which is
inconsistent with the Einstein equations except for the recollapsing
positively-curved FRW metrics. Even then the condition $a(\tau )=a(\tau +nT)$
is only satisfied at one catastrophic time, namely at the singularity, so
identifying time there is overkill. More generally, Geroch \cite{ger} has
shown that if spacetime is globally hyperbolic, so that initial data on a
spacelike slice determine the entire global structure of spacetime uniquely
and completely, then the global topology of spacetime must be $\Sigma \times
R$, where $\Sigma $ is the topology of any Cauchy hypersurface. If one
departs from exact FRW models then the complexities of entropy production,
gravitational clustering and black hole formation will inevitably make any
final singularity different in structure than the initial 
one\footnote{In the non-general relativistic context we are 
familiar with the time
recurrence 'paradox' of Poincar\'{e}, but Tipler \cite{{tipler1},{tipler2}}
has
shown that this recurrence does not occur in general relativity when gravity
is attractive ($\rho +3p>0$). Two states of a generic 
positively curved universe cannot
be identical or even arbitrarily close. For a shorter proof see also 
\cite{gall}. The initial and final singularity could be interpreted as the
recurrent state but we do not expect them to be identical}.  
Therefore,
when looking for topology in CMB maps, there is no obviously natural way to
include compact time or Lorenz transformations.

If we suspend the restriction to expanding FRW models, we can consider
compact time in a hypothetical universe. Since space and time might be on an
equal footing, we could wonder if time is topologically identified at the
big bang or its quantum analogue. While it is conceivable that topological
conditions are set across very small spatial sections initially, it is
harder to imagine how topological conditions could be fixed across large
time intervals. The topology of the universe will presumably be set by
quantum gravity across a Planck scale and similarly across a Planck timespan
because of the fundamental ambiguities in the definition and distinction of
space and time in the quantum gravity era. A quantum initial state is then
blown up by the expansion of the cosmos, so that three space dimensions
become very large (although further space dimensions may have remained quite
small). There is no known reason why only one time dimension exists (or is
perceived) and string theories only pick out special dimensions of
spacetime, never of space and time separately,  but there are persuasive
anthropic reasons why $(3+1)$-dimensional spacetimes are needed for
observers to exist \cite{{bt},{tt}}. 
Would a topological cyclic time always expand
so that any temporal periodicity would exceed cosmological timescales?
If entropy always increases, one might ask how the entire universe in the
future could approach a lower entropy state identical to the universe in the
past?

While it is not obvious how to generate a large compact time direction from
a big bang, we can for the sake of argument assume a universe that has a
compact time direction ab initio. The full isometry group of a 
$(3+1)$-dimensional Minkowski spacetime includes Lorenz boosts as well as
translations and rotations. There are two distinct ways in which to involve
time in the compactification. First, we could identify time under a simple
translation so that $\tau =\tau +nT,$ where $n$ is an integer. This is a
simple closing of time into a circle. Another possibility would be to
identify spacetime points under boosts so that 
\begin{equation}
x\rightarrow \Lambda x,
\end{equation}
which leads to 
\begin{equation}
\pmatrix{\tau\cr x}\rightarrow \pmatrix{\tau\cosh nb+x\sinh nb \cr x\sinh nb
+ \tau \cosh nb}\rightarrow \pmatrix{\tau^\prime\cr x^\prime},
\end{equation}
where $n$ is an integer. Twins would now encounter genuine factual conflicts
since $(\tau ,x)$ is identified with $(\tau ^{\prime },x^{\prime }).$ They
must be the same age but they have no uniquely definable ages. The result is
the Misner spacetime which featured prominently in discussions of the
Chronology Projection Conjecture \cite{{misner},{gl}}. 
Both the circular time universe and the Misner spacetime support closed
timelike loops.

With circular time, if the spatial sections are infinite so that spacetime
is topologically like a cylinder, then only observers at rest in the
preferred frame will see closed timelike loops. Any observer $x^{\prime }$
moving at relative velocity $v$ with respect to $x$ will never be able to
execute a closed timelike circuit in spacetime. However, if space is
identified under a translation, as with the hypertorus, then an observer
could follow a closed timelike curve by executing a rational number of
windings around space relative to his windings in time. Technically, this
would be a set of measure zero amongst all relative windings but the winding
ratio may be necessarily rational because of the physical requirements of
finite physical resolution.

One could argue that there could be no kill-your-grandmother paradox since
topologically identified events are identical. If an event happened once, it
would happen in exactly the same way each time the spacetime point was
revisited. No free will would be possible. We couldn't choose to commit the
murderous act. We couldn't age. We couldn't permanently change anything. We
couldn't reorganize things that started in a disordered state. We could
transfer no information.
And for all intents and purposes time would stand
still.

\section*{Acknowledgements}

JL is supported by a PPARC Advanced Fellowship and an award from NESTA.


\begin{references}
\bibitem{bt} J.D. Barrow and F.J. Tipler, 
{\it The Anthropic Cosmological Principle},
(Oxford University Press: Oxford (1986)

\bibitem{bl} J.D. Barrow and J.Levin, Phys. Rev. A {\bf 63} (2001).

\bibitem{weeks} J. Weeks, unpublished.

\bibitem{peters} P.C. Peters, Am. J. Phys. {\bf 51} (1983) 791; P.C. Peters,
Am. J. Phys. {\bf 54} (1986) 334.

\bibitem{lh} J.R. Lucas and P.E. Hodgson, {\it Spacetime and Electromagnetism}
 (Oxford University Press: 1990) pp 76-83.

\bibitem{lum} M. Lachieze-Rey and J. -P. Luminet, Phys. Rept. {\bf 25} 136
(1995).

\bibitem{cqg} The entire issue of Class. Quant. Grav. {\bf 15} 2589 (1998),
G. Starkman editor.

\bibitem{review} J. Levin, Phys. Rept. {\bf 365} 251  (2002).

\bibitem{Thurston} W.P. Thurston, ``Three-dimensional geometry and
topology'' (Ed. S. Levy, Princeton University Press, Princeton, N.J.
1997).

\bibitem{wmap} D. Spergel et. al., astro-ph/0302209.

\bibitem{angelica} M. Tegmark and A. de Oliviera-Costa, astro-ph/0302496.

\bibitem{efstath} G. Efstathiou, astro-ph/0303127.

\bibitem{css} N.J. Cornish, D.N. Spergel and G. Starkman, Phys. Rev. D 
{\bf 57} (1998) 5982.

\bibitem{pat} J. Levin, E. Scannapieco, G. Gasperis, J. Silk and J. D.
Barrow, {\em Phys. Rev. } {\ D} {\bf 58} (1998) article 123006.

\bibitem{kod1} J.D. Barrow and H. Kodama, Class. Q. Grav. 18, 1753 (2001).

\bibitem{kod2} J.D. Barrow and H. Kodama, Int. J. Mod. Phys. D
{\bf 10} 785 (2001).

\bibitem{king} D.H. King, in Einstein Studies vol 6: Mach's principle: 
From Newton's
Bucket to Quantum Gravity, pp. 237-248, Birkh\"auser, Boston (1995).

\bibitem{ger} R. Geroch, J. Math. Phys. {\bf 11} 437 (1970)

\bibitem{tipler1} F.J. Tipler, Nature {\bf 280} 203 (1979).

\bibitem{tipler2} F.J. Tipler, in Essays in General Relativity: 
A Festschrift for
Abraham H. Taub, ed F.J. Tipler, pp. 21-37, Academic Press, NY.

\bibitem{gall} G.J. Galloway, Comm. Math, Phys. {\bf 96} 423 (1984).

\bibitem{tt} M. Tegmark, Ann. Phys. {\bf 270} 1 (1998).

\bibitem{misner} C.W. Misner, in {\it Relativity Theory and Astrophysics I:
Relativity and Cosmology}, edited by J. Ehlers, Lectures in Applied
Mathematics,  8 (American Mathematical Society, Providence, 1967), p.160.

\bibitem{gl} J.R. Gott and L-X. Li, Phys. Rev. D {\bf 58} 023501 (1998).
\end{references}
\end{document}